\documentclass[%
 reprint,
 amsmath,amssymb,
 aps,
]{revtex4-1}

\usepackage{graphicx}
\usepackage{dcolumn}
\usepackage{bm}
\usepackage{wrapfig}

\newcommand{\beq}{\begin{equation}}
\newcommand{\eeq}{\end{equation}}
\newcommand{\beqa}{\begin{eqnarray}}
\newcommand{\eeqa}{\end{eqnarray}}

\newcommand{\kBT}{\mbox{$k_{\rm B}T$}}

\begin{document}


\title{
Relationship Between Macrostep Height and Surface Velocity for a Reaction-Limited Crystal Growth Process
}

\author{Noriko Akutsu}
\email{nori3@phys.osakac.ac.jp, nori@phys.osakac.ac.jp}
\affiliation{%
Faculty of Engineering, Osaka Electro-Communication University, Hatsu-cho, Neyagawa, Osaka 572-8530, Japan
}%




\date{\today}

\begin{abstract}
\begin{wrapfigure}{r}{7.5cm}
\hspace*{-4.5cm} 
\includegraphics[width=8cm]{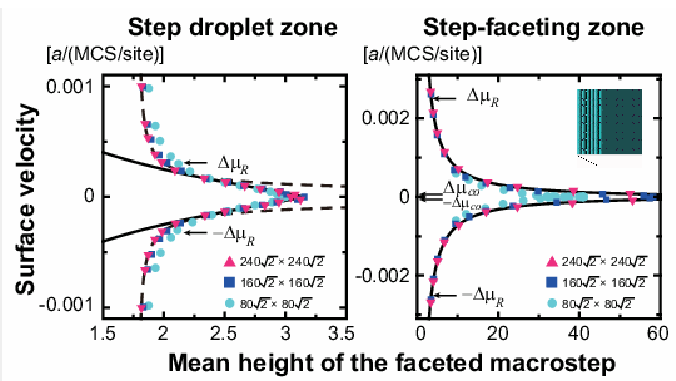}
\end{wrapfigure}
This work examined the effect of macrostep height on the growth velocity of a vicinal surface during reaction- (interface-) limited crystal growth under non-equilibrium steady state conditions. The Monte Carlo method was employed, based on a  restricted solid-on-solid (RSOS) model with point-contact-type step-step attraction (termed the p-RSOS model).
Although this is a simple lattice model, the model surface shows a variety of distinctive configurations depending on the temperature and the driving force for crystal growth.
The results demonstrate that the surface velocity decreases as the height of the faceted macrostep increases.  
In addition, the significant variation in surface velocity recently reported by Onuma {\it et al.} in a study based on 4H-SiC was reproduced.
This work also shows that the terrace slope, elementary step velocity and elementary step kinetic coefficient are all affected by the faceted macrostep height.
\begin{description}
\item[PACS numbers]
81.10.Aj 64.60.Q- 82.60.Nh 68.35.Md 02.70.Uu 81.10.Dn 68.35.Ct 05.70.Np 
\end{description}
\end{abstract}


\maketitle


\section{Introduction}
The production of high-quality SiC crystals is an important prerequisite for the production of advanced power devices with low power consumption rates.  
Although self-organized faceted macrosteps are known to lower the quality of crystalline SiC \cite{mitani}, dislocations penetrating the crystal have been shown to end at the side surfaces of macrosteps \cite{usui,miyake,harada}. 
Hence, the intentional introduction of macrosteps can effectively decrease the dislocation density in a SiC crystal. 
In addition, controlling the self-assembly and disassembly of faceted macrosteps is an important aspect of fabricating semiconductor crystals.

There have been numerous studies concerning step bunching during diffusion limited crystal growth \cite{pimpinelli,  barabasi,misba10, krzy14, krzy17}.
As an example, Chernov \cite{chernov61} and co-workers \cite{nishinaga89, nishinaga93} performed detailed studies of the diffusion field around macrosteps or trains of macrosteps during steady-state growth in which volume diffusion was the rate-determining process. 
This same prior work also generated mathematical expressions for the velocity of macrostep advancement $V_m$ under these conditions, {\it i. e.} $V_m \propto 1/h_m$ where $h_m$ is the height of the macrostep \cite{chernov61}.
The essential point to derive this equation is the {\it mass conservation} during the diffusion limited crystal growth.
Recently, Onuma {\it et al.} \cite{onuma17} determined the relationship between the macrostep velocities and heights in 4H-SiC using {\it in situ} confocal laser scanning microscopy and found that, when employing a Si--Ni flux, the experimental data could be explained by Chernov's equation \cite{chernov61}.
Thus, they concluded that the rate-determining process in 4H-SiC crystal growth is volume diffusion.
In contrast, when Al was added to the Si--Ni flux, the  data were too scattered to be explained by the equation, and so  the rate was thought to have been determined by the interfacial reactions.

Despite this prior work, reaction- (that is, interface-) limited growth has not been studied sufficiently.
In addition, such interfacial (or surface) reactions are always accompanied by volume diffusion or surface diffusion, and so it is difficult to separate the effects of interfacial reactions from those of diffusion in an actual system. 
For this reason, the present work employed computational experiments involving extreme conditions, using the Monte Carlo method.

We have been studying the self-assembly/disassembly of macrosteps on a theoretical basis, neglecting the effects of the surface and volume diffusion.
The main driving force for the assembly of steps in such cases is 
anomalous surface tension at low temperatures.
The anomaly in surface tension is caused microscopically by step-step attraction.
This prior work employed the restricted solid-on-solid (RSOS) model in conjunction with point contact step-step attraction, termed the p-RSOS model \cite{akutsu09, akutsu11, akutsu11JPCM, akutsu12, akutsu14, akutsu16-2}.
Point contact step-step attraction is considered to result from the energy gain associated with the point at which neighboring steps meet.
At such meeting points, the dangling bonds of the neighboring steps overlap to merge to create actual bonds between atoms.
The  resulting energy gain  leads to the step-step attraction.
Although the p-RSOS model involves a simple lattice, the vicinal surface of the model exhibits a variety of surface configurations with respect to the self-assembling/disassembling of macrosteps depending on the temperature and the driving force for crystal growth \cite{akutsu16, akutsu16-3, akutsu17, akutsu18}.

The most important aspect of the p-RSOS model is that it allows reliable calculations of a polar graph of the surface tension (the Wulff figure) and the equilibrium crystal shape (ECS). 
The surface tension is surface free energy per unit normal area.
Since surfaces and steps are low-dimensional objects, thermal fluctuations are so severe that they destroy the ordered phase for the system with the short-range force, generally \cite{mermin}.
The calculation of the surface free energy with the mean-field or the quasi-chemical approximation often leads to wrong results.
Hence, the calculations more precise than the mean field approximation are required.
In the present work, we applied the transfer matrix version of the density matrix renormalization group (DMRG) method \cite{dmrg, dmrg2, dmrg3,pwfrg, pwfrg2, pwfrg3, akutsu98, akutsu06} to calculate the surface tension.
Therefore, the morphological phenomena resulting from the anisotropy of the surface tension can be analyzed  by referring to the calculated Wulff figures.

The aim of this work was to demonstrate the effect of the macrostep height on the surface velocity, the terrace slope, the elementary step velocity, and the elementary step kinetic coefficient in the case of reaction (interface)-limited crystal growth in the non-equilibrium steady state.
This involved simulating the vicinal surface  based on the p-RSOS model and using the Monte Carlo method in the non-conserved system of mass.

\section{Microscopic Model and Calculation Method}

\subsection{The p-RSOS model}


\begin{figure}
\centering
\includegraphics[width=8.0 cm,clip]{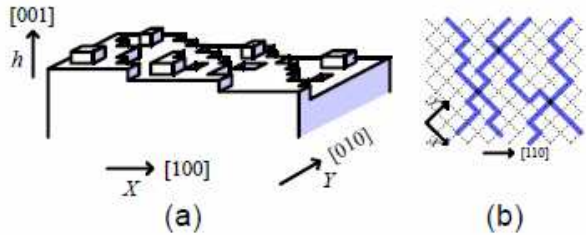}%
\caption{
(a) Perspective view of the RSOS model tilted towards the $\langle 110\rangle$ direction. (b) Overhead view of the RSOS model. The thick blue lines indicate the surface steps.
This figure is taken from [22].
}  
\label{vicinal}
\end{figure}


\begin{figure*}
\centering
\begin{minipage}{11pc}
\includegraphics[width=9pc]{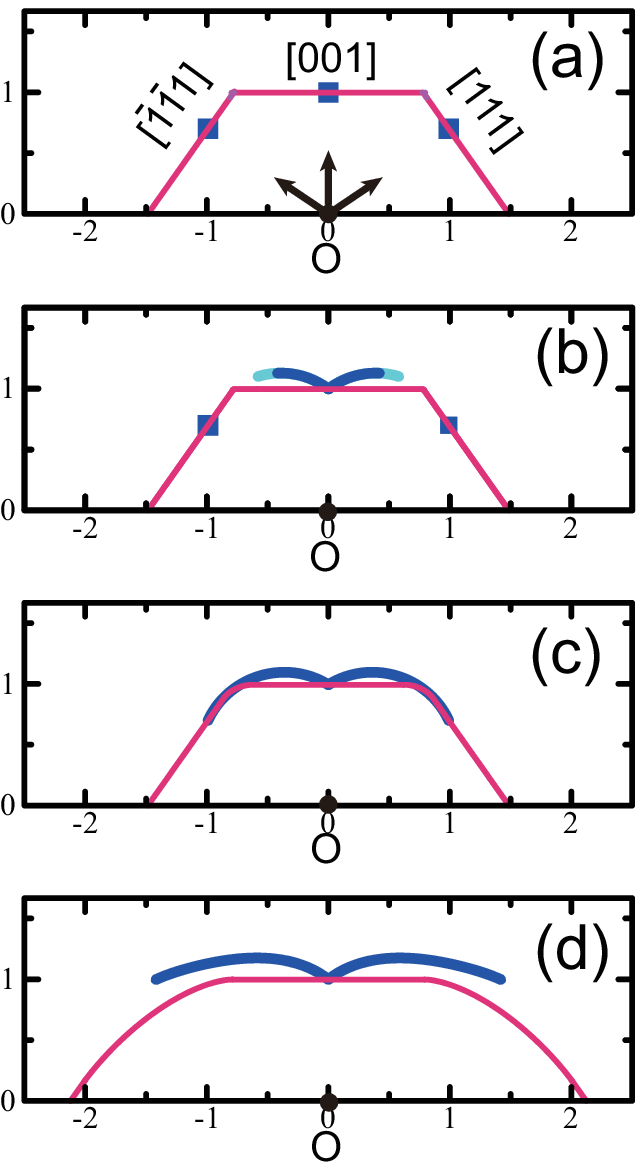}
\caption{\label{fig_wulff}Wulff figures.  
The thick dark, thick light and thin lines indicate polar graphs of the surface tension, polar graphs of the metastable surface tension and ECS profiles, respectively.
$\epsilon_{\rm int}/\epsilon= -0.9.$ (a) $\kBT/\epsilon=0.6.$ (b) $\kBT/\epsilon=0.63.$ (c) $\kBT/\epsilon=0.73.$ (d) Original RSOS model at $\kBT/\epsilon=0.6$ and $\epsilon_{\rm int}=0.$ This figure is taken from [21].} \label{wulff_fig}
\end{minipage}\hspace{2pc}%
\begin{minipage}{24pc}
\includegraphics[width=6.5 cm,clip]{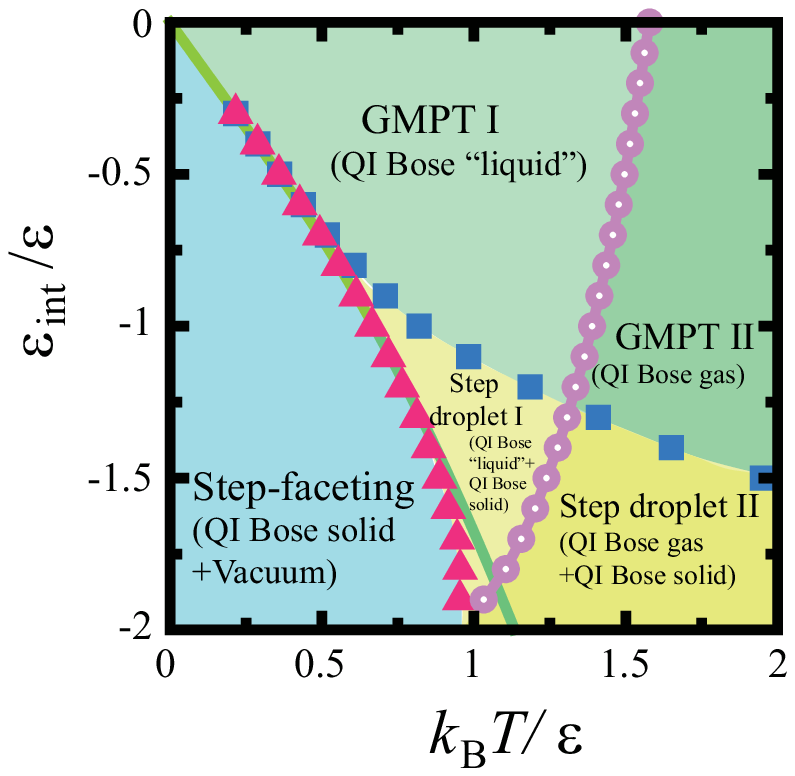}%
\caption{
Faceting diagram \cite{akutsu16}. Red triangles indicate $T_{f,2}$ values.  
Below $T_{f,2}$, the polar graph of the surface tension is discontinuous around the (001) surface \cite{akutsu11JPCM,akutsu12,akutsu16}.
Purple squares indicate $T_{f,1}$ values.
Below $T_{f,1}$, the polar graph of the surface tension is discontinuous around the (111) surface \cite{akutsu11JPCM,akutsu12,akutsu16}.
Pink circles indicate the roughening transition temperatures of the (001) surface. 
All  symbols show values calculated using the PWFRG method.  
As for the ``QI Bose solid'', ``QI Bose liquid'', and ``AI Bose gas'', please refer to the Ref. [20]. 
This figure is taken from [20].
}  
\label{fig_diagram}
\end{minipage}\hspace{2pc}%
\end{figure*}

In this study, the p-RSOS model was employed as the microscopic model (Fig.~\ref{vicinal}) \cite{akutsu09}-\cite{akutsu18}. 
In this approach, the total energy of the (001) surface can be written as
\beqa
&&{\cal H}_{\rm p-RSOS} = {\cal N}\epsilon_{\rm surf}+   \nonumber \\
&&\sum_{n,m} \epsilon 
[ |h(n+1,m)-h(n,m)|  \nonumber \\
&&
+|h(n,m+1)-h(n,m)|]   \nonumber \\
&&
 +\sum_{n,m} \epsilon_{\rm int}[ \delta(|h(n+1,m+1)-h(n,m)|,2)  \nonumber \\
&& 
+\delta(|h(n+1,m-1)-h(n,m)|,2)],   \label{hamil}
\eeqa
where ${\cal N}$ is the total number of lattice points, $\epsilon_{\rm surf}$ is the surface energy per unit cell on the planar (001) surface, $\epsilon$ is the microscopic ledge energy, $\delta(a,b)$ is the Kronecker delta and $\epsilon_{\rm int}$ is the microscopic step-step interaction energy.
$\epsilon_{\rm int}$ contributes to the surface energy at the meeting point of neighboring steps when the height difference in the diagonal direction is $\pm 2$, and is assumed to originate from the energy gain when overlapping dangling bonds at step edges form bonding states.
The summation with respect to $(n,m)$ is performed over all sites on the square lattice and when $\epsilon_{\rm int}$ is negative, the step-step interaction becomes attractive.

It should also be noted that the RSOS model implicitly requires that the height difference between the nearest neighbor sites be restricted to $\{ 0, \pm 1\}$. 
The surface free energy density, surface tension and ECS can be calculated using the transfer matrix version of the DMRG method \cite{dmrg, dmrg2,dmrg3}, termed the product-wave-function renormalization group (PWFRG, or tensor network) method \cite{pwfrg,pwfrg2,pwfrg3,akutsu98,akutsu15book}.

After detailed calculations of the surface free energy density and the surface tension, we found out that the surface undergoes the first order transition with respect to the self-assembling/disassembling of the elementary steps in the vicinal surface (Fig. \ref{wulff_fig}). 
At high temperatures, the vicinal surface exhibits the Gruber-Mullins-Pokrovsky-Talapov (GMPT) universal behavior \cite{akutsu98, gmpt, gmpt2,  akutsu15book} (Fig. \ref{wulff_fig} (c)). 
As the temperature is decreased, the surface tension around the (111) surface becomes discontinuous (Fig. \ref{wulff_fig} (b)) below $T_{f,1}$, while the surface tension around the (001) surface becomes discontinuous (Fig. \ref{wulff_fig} (a)) below $T_{f,2}$.
The faceting diagram shown in Fig. \ref{fig_diagram} summarizes the surface tension discontinuities.

The profile of the faceted macrostep on the vicinal surface at equilibrium is also determined by the surface tension continuity  \cite{akutsu16-3}.
At $T>T_{f,1}$ (GMPT zone), a faceted macrostep does not appear on the vicinal surface.
In contrast, at $T_{f,1} \geq T>T_{f,2}$ (that is, the step-droplet zone), a faceted macrostep with the side surface being the $(111)$ surface coexists with vicinal surfaces having the slope $p_1$. 
Note that vicinal surfaces with the slope $p_1<p (111)$ do not appear, because they are thermodynamically unstable.
At $T<T_{f,2}$ (that is, the step-faceting zone), the vicinal surface consists of only the $(001)$ and the $(111)$ surfaces.


\subsection{Monte Carlo method}

The Monte Carlo method was adopted together with the Metropolis algorithm for the non-conserved systems to study the non-equilibrium steady state.
The microscopic surface energy for a fixed number of steps, $N_{\rm step}$, was determined using the equation
\beq
{\mathcal H}_{\rm non eq}= {\mathcal H}_{\rm p-RSOS} 
- \Delta \mu \sum_{n,m} h(n,m,t), \label{hamil_neq}
\eeq
where $t$ is the time in units of Monte Carlo steps per site (MCS/site) and $\Delta \mu$ is the driving force for the crystal growth (that is, the chemical potential difference between the bulk crystal and the ambient phase).
When $\Delta \mu>0$, the crystal grows, whereas when $\Delta \mu<0$, the crystal recedes (evaporates, dissociates or melts).

The site on the surface for an Monte Carlo event is chosen randomly.
Addition or removal of an growth unit is determined with the probability 0.5.
The energies of the surface before or after of the event $E_i$ or $E_f$ are calculated based on Eq. (\ref{hamil_neq}).
If $E_f-E_i \leq 0$, the event occurs with the probability 1; whereas if $E_f-E_i > 0$, the event occurs with the probability $\exp[ - (E_f-E_i)/\kBT]$.


\begin{figure*}
\centering
\includegraphics[width=16.0 cm,clip]{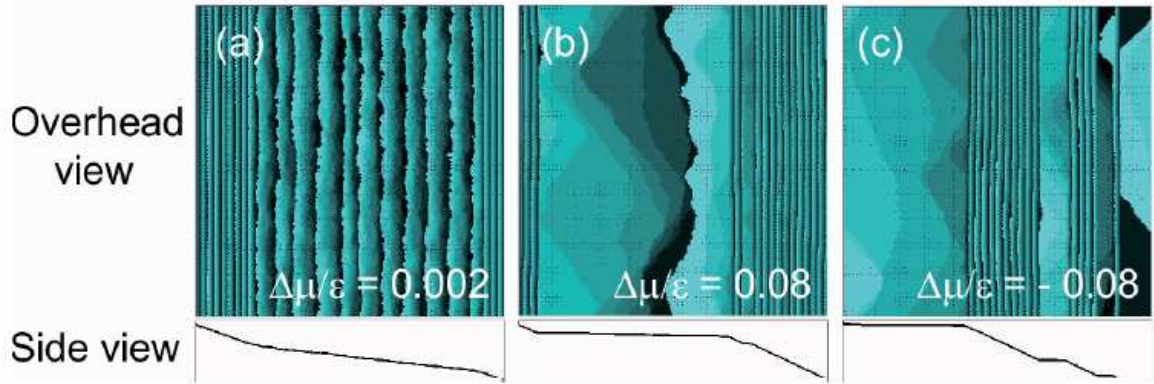}%
\caption{
Images of the vicinal surface generated using the Monte Carlo method at $4 \times 10^8$ MCS/site.
The brightness in each image is proportional to the height of the surface, with 10 gradations.
(a) A typical surface in the step-droplet zone ($\kBT/\epsilon=0.63$). $N_{\rm step}=240$.   
(b) and (c) Typical surfaces in the step-faceting zone ($\kBT/\epsilon=0.4$) in the case of growth and recession (evaporation, dissolution), respectively. $N_{\rm step}=180$.
Size: $240 \sqrt{2} \times 240 \sqrt{2}$.
  $\epsilon_{\rm int}/\epsilon=-0.9$.
}  
\label{surfdat}
\end{figure*}

Periodic boundary conditions were required in the direction of the mean step-running.
In the direction normal to the mean step-running, the surface heights of the lowest side line were connected to the surface heights at the topmost line by adding the number of elementary steps.   
The initial surface configuration involved the preparation of a parallel train of steps and a single macrostep.

It should be noted that the external parameters are the microscopic ledge energy $\epsilon$, microscopic step-step attraction $\epsilon_{\rm int}$ ($<0$), temperature $T$, total number of elementary steps $N_{\rm step}$,  linear system size $L$, and driving force for the crystal growth $\Delta \mu$.

\subsection{Macrostep height}
Images of the simulated surfaces at $4\times 10^8$ MCS/site are presented in Fig. \ref{surfdat}.
The average macrostep height was obtained using the equation 
\beqa
\langle n \rangle &=&\sum_{\tilde{y}}  \sum_{\tilde{x}}|n_{\tilde{x}}(\tilde{y})|/  [ \sum_{\tilde{y}} n_{\rm step}(\tilde{y}) ]  \nonumber \\
&\approx& N_{\rm step}/{\langle n_{\rm step} \rangle },
\eeqa
where $\tilde{x}$ is the $\langle 110 \rangle$ direction (normal to the mean step-running direction), $\tilde{y}$ is the $\langle \bar{1}10 \rangle$ direction (along the mean step-running direction) and $n_{\rm step}$ is the number of merged steps.
When determining the average value, the data over the first $2 \times 10^8$ MCS/site were discarded and the values over the subsequent $2 \times 10^8$ MCS/site were averaged.
The surface velocity $V$ was estimated by finding the average surface height $\bar{h}(t)$, using the formula 
\beq
\bar{h}(t)= (1/ {\mathcal N})\sum_{n,m}h(n,m).
\eeq
Here, $V$ is defined as
\beq
V=[\bar{h}(t_{\rm max})-\bar{h}(t_0)]/(t_{\rm max}-t_0),
\eeq
where $t_0$ and $t_{\rm max}$ are $2\times10^8$ MCS/site and $4\times10^8$ MCS/site, respectively.


\section{Results and discussion}
\subsection{\label{sec_vn} Macrostep height dependence of the surface velocity}

\begin{figure*}
\centering
\includegraphics[width=15.0 cm,clip]{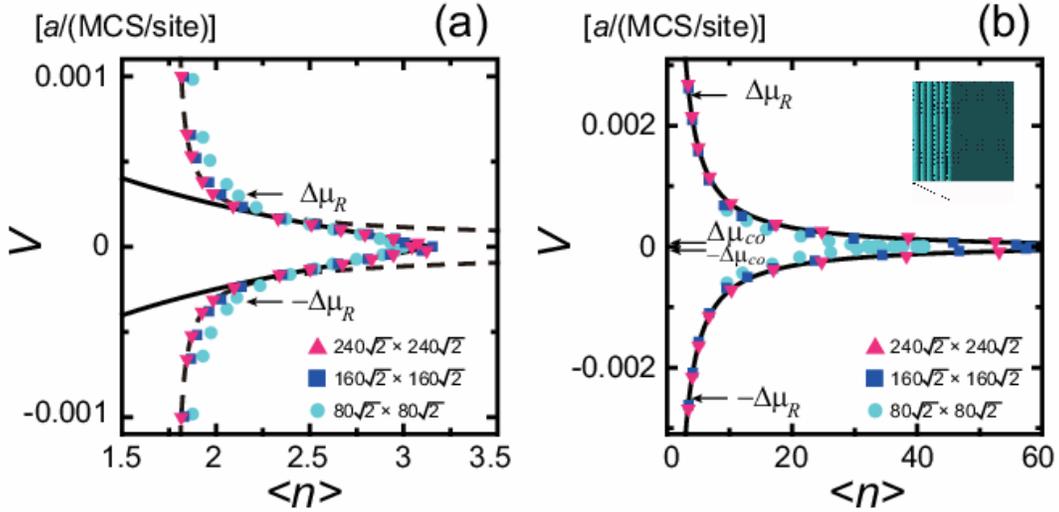}%
\caption{
Macrostep height dependence of the surface velocity.
$V>0$ indicates crystal growth, whereas $V<0$ indicates dissolution (or etching) of the crystal.
$\epsilon_{\rm int}/\epsilon = -0.9$.
(a) In the step-droplet zone.
$\kBT/\epsilon= 0.63$.
Solid lines: Eq. (\ref{eq_v-nt063}).
Broken lines: Eq. (\ref{eq_VndmuR}).
Here, the macrostep velocity is proportional to the surface velocity.
(b) In the step-faceting zone.
$\kBT/\epsilon= 0.4$.
Solid lines: Eqs. (\ref{eq_k})--(\ref{eq_n}).
The inset in (b) shows the initial configuration of the surface.
The macrostep velocity is almost zero.
}  
\label{V-n}
\end{figure*}

\begin{center}
\begin{table}
\caption{\label{table1} Characteristic driving forces ($L/(\sqrt{2}a)=240$, $a=1.$) \cite{akutsu18}}
\centering
\begin{tabular}{clcl}
\hline
Symbol& value$/\epsilon$ & $\kBT/\epsilon$ & description\\
\hline
\vspace{-0.7em}&&\\
$\Delta \mu_R(L)$ & $0.121\pm0.005$ & 0.4 &  {Crossover point between the }\\
& $0.005 \pm 0.002 $ & 0.63 & step-detachment mode and \\
&&& the kinetically roughened mode.\\[2mm]
$\Delta \mu_{co}(L)$ & $0.050\pm 0.007$ &0.4 & Crossover point from the 2D\\
&&&  nucleation mode to the successive \\
&&&  step-detachment mode.\\[2mm] 
$\Delta \mu_{f}(L)$ & $0.023\pm 0.007$ &0.4 & Freezing point\\[2mm] 
$\Delta \mu_y (L)$ & $0.018 \pm 0.006$& 0.4 & Yielding point  \\
\hline
\end{tabular}
\end{table}
\end{center}

The effects of height $\langle n \rangle$ on the surface velocity are shown in Fig. \ref{V-n}. 
These data demonstrate that surface velocity $V$ decreases as the mean size of the macrostep, $\langle n \rangle$, increases.
It should be noted that both $V$ and $\langle n \rangle$ are functions of driving force for the crystal growth $\Delta \mu$ in the non-equilibrium steady state \cite{akutsu18}.
According to our previous work \cite{akutsu18}, there are several characteristic driving forces, which are summarized in Table \ref{table1}.

\begin{figure}
\centering
\includegraphics[width=6.0 cm,clip]{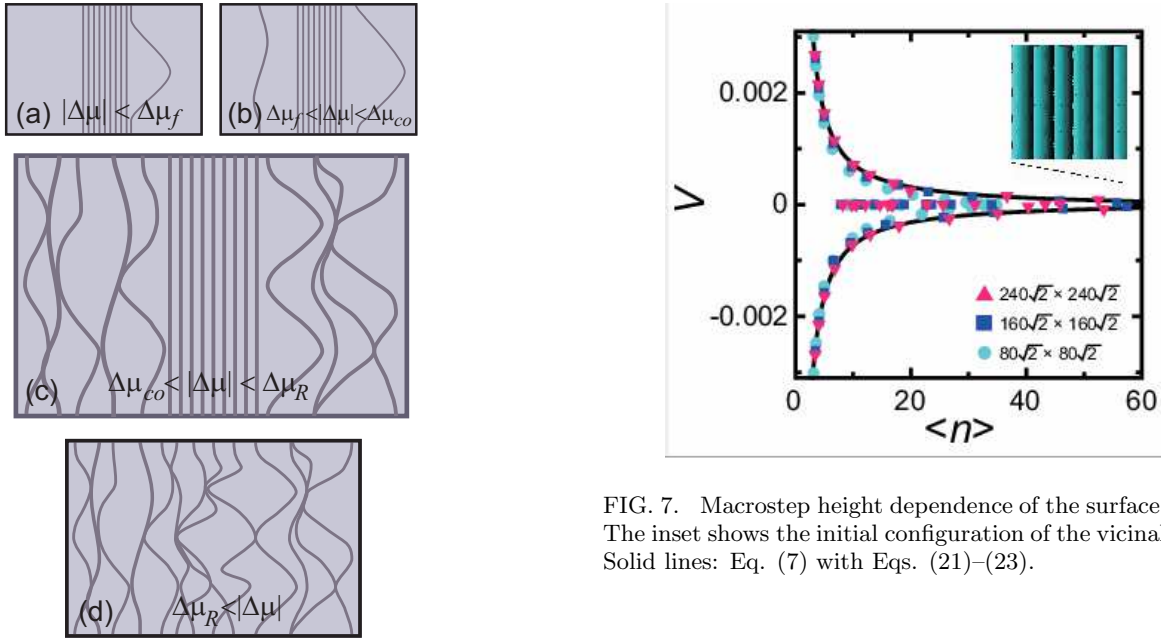}%
\caption{
Characteristic step configurations of a vicinal surface. 
The left side of the surface is higher than the right side.
}  
\label{fig_characteristics}
\end{figure}

In the step-droplet zone, for $|\Delta \mu| < \Delta \mu_R$, the vicinal surface has the structure shown in Fig. \ref{surfdat} (a), {\it i. e.} a faceted macrostep with a (111) side surface in conjunction  ``terrace'' surfaces having the slope $p_1$. 
The characteristics of this structure are provided in Fig. \ref{fig_characteristics} (c). 
In contrast, in the case of $|\Delta \mu| > \Delta \mu_R$, the vicinal surface is kinetically roughened and a macrostep is not formed.
The characteristics of this structure are shown in Fig. \ref{fig_characteristics} (d). 
Although there is no macrostep, the elementary steps merge locally to form the {\it faceted local macrosteps}. 
In addition, $\langle n \rangle$ and $V$ exhibit power law behaviors \cite{akutsu17}.

\begin{figure}
\centering
\includegraphics[width=7.0 cm,clip]{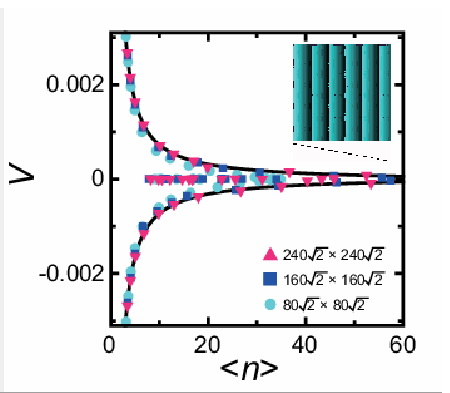}%
\caption{
Macrostep height dependence of the surface velocity.
The inset shows the initial configuration of the vicinal surface.
Solid lines: Eq. (\ref{eq_Vdef}) with Eqs. (\ref{eq_k})--(\ref{eq_n}). 
}  
\label{fig_t04para}
\end{figure}

In the step-faceting zone, the change in dynamics is more complex than in the step-droplet zone, as there are characteristic driving forces near equilibrium in addition to $\Delta \mu_R$.
At equilibrium, only the (001) and (111) surfaces are thermodynamically stable, and the (001) surface forms the terrace, while the (111) surface forms the side surface of the macrostep.
In the event that $|\Delta \mu| < \Delta \mu_{f}$, the surface does not grow, because there is a lack of nucleation growth, meaning that the mean waiting time for the formation of a single nucleus exceeds the observation time ($4 \times 10^8$ MCS/site in this work).  
In the case that a 2D island is formed at the edge of the macrostep as a result of thermal fluctuations, the island shrinks because the elementary step that surrounds the island recedes back to the macrostep due to the step tension  (Fig. \ref{fig_characteristics} (a)).
In addition, since the vicinal surface freezes for $|\Delta \mu| < \Delta \mu_{f}$, the height of the faceted macrostep is largely determined by the initial configuration.
As a result, the relationship between $\langle n \rangle$ and $V$ also depends on the initial configuration, and so the $\langle n \rangle$--$V$ plot in Fig. \ref{fig_t04para}  exhibits significant scatter in the region between the solid lines. 

\begin{figure}
\centering
\includegraphics[width=7 cm,clip]{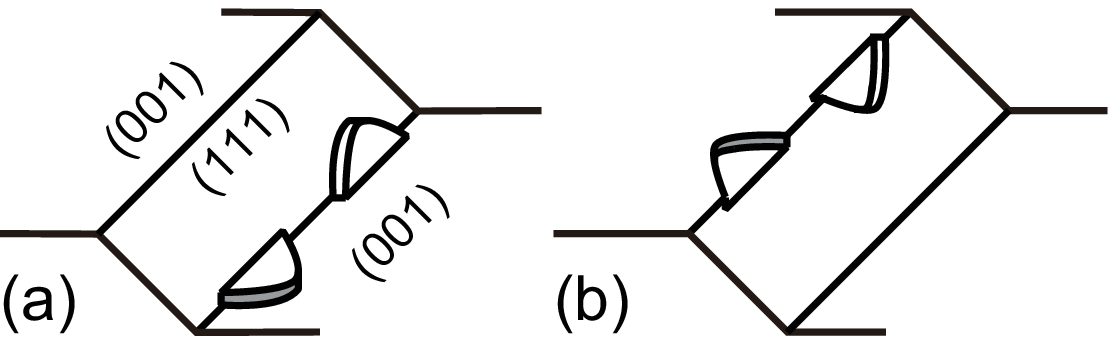}%
\caption{
Schematic illustration of 2D nucleation. 
(a) Growth.
(b) Recession (evaporation or dissolution).
}  
\label{fig_nucleations}
\end{figure}
 
In the case of $ \Delta \mu_{f}< |\Delta \mu| < \Delta \mu_{co}$, the crystal growth proceeds via 2D single nucleation (Fig. \ref{fig_characteristics} (b)).
If $\Delta \mu >0$, then the crystal grows and nucleation occurs at the lower edge of the macrostep while growth occurs on both the (001) and  (111) surfaces (Fig. \ref{fig_nucleations} (a)).
Conversely, if $\Delta \mu <0$, then the crystal evaporates or dissolves. 
A ``negative'' 2D island, which is the cluster of holes on a surface,  (Fig. \ref{surfdat} (c)) is formed as a nucleus at the upper edge of the macrostep, and this negative island increases in size (Fig. \ref{fig_nucleations} (b)).

If $ \Delta \mu_{co}< |\Delta \mu| < \Delta \mu_{R}$, then the crystal grows via 2D multi nucleation  (Fig. \ref{fig_characteristics} (c)).
If $\Delta \mu_{R} \leq |\Delta \mu|$, then the surface is kinetically roughened, and both $V$ and $\langle n \rangle$ demonstrate power law behavior with respect to $|\Delta \mu|$.
Although the relationship between $V$ and $\langle n \rangle$ in the step-faceting zone is similar to that in the step-droplet zone, the details of the relationships are different. 
To understand this phenomenon on the microscopic scale, these Monte Carlo results are further analyzed in the following subsections.

\subsection{Step attachment/detachment model \label{sec_attach}}

In the case that two surfaces coexist, elementary steps will attach to or detach from the edge of the faceted side surface.  
The effect of $\Delta \mu$ on the vicinal surface morphology for $|\Delta \mu| < \Delta \mu_{R}$ was assessed by considering a step--attachment--detachment model as $\langle n \rangle$ evolved over time \cite{akutsu17}. This was calculated as
\beq
\frac{\partial \langle n \rangle}{\partial t}= n_+ - n_-, \label{eq_dn/dt}
\eeq
where $n_+$ is the rate at which the elementary steps catch up to a macrostep and $n_-$ is the rate at which the elementary steps detach from a macrostep.
When $n_+<n_-$, the macrostep dissociates (as is the case for $\Delta \mu_{R}<|\Delta \mu|$), 
whereas when $n_+>n_-$ (the case for $\Delta \mu_{f}<|\Delta \mu| < \Delta \mu_{co}$), $\langle n \rangle$ increases up to $N_{\rm step}$.
In this case, $n_-$ limits the growth/recession rate of the surface.
Under steady-state conditions, $n_+ = n_- = V/a$, where $a$ is the height of the elementary step.
Thus, if $|\Delta \mu| < \Delta \mu_{R}$, then $n_+$ can be expressed as $n_+= \rho_1 v_1$, where $\rho_1$ and $v_1$ are the step density and the step velocity of the elementary steps, respectively.


Based on the step--attachment--detachment model, surface velocity $V$ can be expressed using more microscopic quantities, as \cite{akutsu17,akutsu18}
\beq
V= p_1 v_1, \label{eq_Vdef} 
\eeq
where $p_1$ is the slope of the terrace surface and $v_1$ is the mean step velocity of the elementary steps. 
Since total number of elementary steps $N_{\rm step}$ is conserved, the slope $p_1$ can be expressed using $\langle n \rangle$ as 
\beqa
p_1&=& \frac{\sqrt{2} \bar{p}(\frac{1}{\langle n \rangle}-\frac{N_m}{N_{\rm step}})}{\sqrt{2}- \bar{p} + \bar{p} (\frac{1}{\langle n \rangle}-\frac{N_m}{N_{\rm step}})},\nonumber \\
&& \bar{p}=N_{\rm step}a/L, \label{eq_p1def}
\eeqa
where $\bar{p}$ is the mean slope of the vicinal surface, $N_m$ is the number of macrosteps and $a$ is the height of the elementary steps.
It should be noted that $p_1=0$ for 
\beq
\langle n \rangle =  N_{\rm step}/N_m = n^* .   \label{n-p1zero}
\eeq
In the case of $\langle n \rangle \sim n^*$, $p_1$ is approximately $\{\sqrt{2} \bar{p} / (\sqrt{2} - \bar{p})\}(1/\langle n \rangle -1/n^*)$.
In addition, if $\langle n \rangle << n^*$, then $1/n^*$ becomes negligible, such that $p_1 \approx \sqrt{2} \bar{p} / \{(\sqrt{2} - \bar{p})\langle n \rangle +  \bar{p} \}$.

%
%


Mean step velocity $v_1$ can be obtained on the basis of $V$ and $\langle n \rangle$ together with Eqs. (\ref{eq_Vdef}) and (\ref{eq_p1def}) as
\beqa
v_1&=&V/p_1  \nonumber \\
&=& V \frac{\sqrt{2}- \bar{p} + \bar{p} (\frac{1}{\langle n \rangle}-\frac{1}{n^*})}{\sqrt{2} \bar{p}(\frac{1}{\langle n \rangle}-\frac{1}{n^*})}.\label{eq_v1def}
\eeqa 
The kinetic coefficient $k= v_1/\Delta \mu$ is also calculated using $V$ and $\langle n \rangle$, as
\beqa
k&=& v_1/\Delta \mu = V/(\Delta \mu \   p_1)   \nonumber \\
&=& \frac{V [\sqrt{2}- \bar{p} + \bar{p} (\frac{1}{\langle n \rangle}-\frac{1}{n^*})]}{\sqrt{2} \Delta \mu \ \bar{p}(\frac{1}{\langle n \rangle}-\frac{1}{n^*})}. \label{eq_kdef}
\eeqa 

Because $\langle n \rangle$ depends on $\Delta \mu$ in the non-equilibrium steady state, a large $\langle n \rangle$ value is associated with $|\Delta \mu| < \Delta \mu_y \approx \Delta \mu_f$ and $p_1 \approx 0$.
This result indicates that the $v_1$ and $k$ terms in Eqs. (\ref{eq_v1def}) and (\ref{eq_kdef}) are divergent near equilibrium (Eq. (\ref{n-p1zero})).
However, $V$ is also affected by $\Delta \mu$ and decreases to zero near equilibrium.
Whether $v_1$ and $k$ are divergent or not near equilibrium depends on the rate of decreases of $V$ and $p_1$.

\subsection{Step-droplet zone}


\begin{figure}
\centering
\includegraphics[width=7.5 cm,clip]{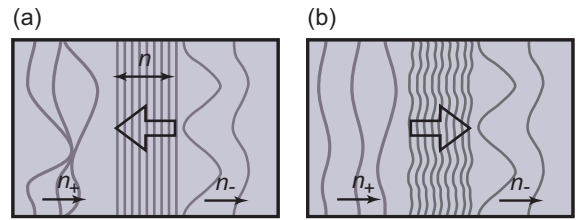}%
\caption{
Schematic illustration of the relationship between macrostep motion and elementary step motion.
The height on the left side is higher than the height on the right.
(a) Faceted macrostep. Typical example of reaction (interface)-limited growth.
(b) Macrostep with a rough side surface. Typical example of diffusion-limited growth.
}  
\label{macrostep_motion}
\end{figure}


\begin{figure*}
\centering
\includegraphics[width=12.0 cm,clip]{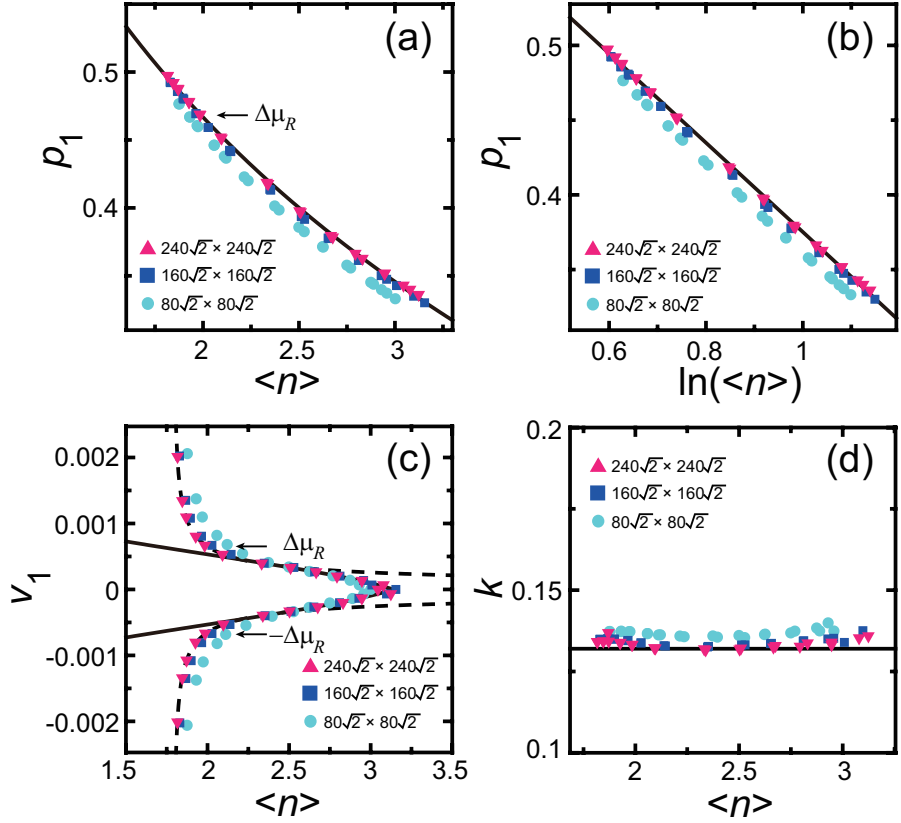}%
\caption{
(a) Macrostep height $\langle n \rangle$ dependence of the surface slope, $p_1$.
The symbols indicate $p_1$ values obtained from Eq. (\ref{eq_p1def}) based on the Monte Carlo data.
Solid line: Eq. (\ref{eq_p1lnn}).
(b) $\ln p_1$ {\it v.s.} $\ln \langle n \rangle$.
Solid line: Eq. (\ref{eq_p1lnn}).
(c) $\langle n \rangle$ dependence of the mean velocity of the elementary steps, $v_1$.
The symbols indicate values obtained from Eq. (\ref{eq_v1def}) based on the Monte Carlo data.
Solid lines: Eq. (\ref{eq_v1-nt063}).
Broken lines: Eq. (\ref{eq_v1ndmuR}).
(d) $\langle n \rangle$ dependence of kinetic coefficient $k$.
The symbols indicate values obtained from Eq. (\ref{eq_kdef}) based on the Monte Carlo data.
Solid line: $k=0.132$.
$\kBT/\epsilon=0.63$.
$\epsilon_{\rm int}/\epsilon=-0.9$. 
$N_{\rm step} = 240$.
}  
\label{fig_p1-vstept063}
\end{figure*}

In the step-droplet zone, the vicinal surface for $|\Delta \mu| < \Delta \mu_R$ has a configuration similar to that in Fig. \ref{fig_characteristics} (c).
It is interesting to observe that the macrostep moves in the opposite direction compared to that of the elementary steps (Fig. \ref{macrostep_motion} (a)). 
This occurs because the side of the faceted macrostep is a (111) smooth surface, such that the number of kinks on this surface is extremely low.
In addition, the vicinal surface grows via the step detachment/attachment mode, and so the effect of $\langle n \rangle$ on the surface velocity $V$ is the same as the effect on the macrostep velocity $V_m$. 

In contrast, in the case of step bunching caused by diffusion-limited growth, the side surface of the bunched step is rough and there are many kinks on the side surface.
Consequently, the center of the bunched step side surface moves in the same direction as the elementary steps (Fig. \ref{macrostep_motion} (b)). These results show that the rate-limiting process can be determined if the growth directions of the macrostep and the elementary steps are identified.
Specifically, when the growth direction of the macrostep is the same as that of the elementary steps, the crystal growth is diffusion-limited, while if the opposite is true, then the crystal growth is reaction (interface)-limited.  

The slope value, $p_1$, can be calculated using Eq. (\ref{eq_p1def}) based on the $\langle n \rangle$ value together with $N_m=2$ and $\bar{p}=\sqrt{2}/2$.
The resulting $p_1$ values are shown in Fig. \ref{fig_p1-vstept063} (a), which demonstrates that $p_1$ decreases as $\langle n \rangle$ increases.
From Fig. \ref{fig_p1-vstept063} (b), we can see that $p_1$ can also be expressed as
\beq
p_1\approx d-f \ln [\langle n \rangle],\label{eq_p1lnn}
\eeq
where
$d=0.674$, and $f= 0.299$. 
The line calculated using Eq. (\ref{eq_p1lnn}) is indicated by the solid lines in Figs. \ref{fig_p1-vstept063} (a) and (b).

The mean velocity of the elementary steps, $v_1$, can be obtained from Eq. (\ref{eq_v1def}) using the values for $V$ and $\langle n \rangle$, and the resulting $v_1$ values are indicated by the symbols in Fig. \ref{fig_p1-vstept063} (c).
These data demonstrate that the absolute value of $v_1$ decreases as $\langle n \rangle$ increases.
It is interesting to note that the kinetic coefficient of the elementary steps, $k$, calculated using Eq. (\ref{eq_kdef}) (Fig. \ref{fig_p1-vstept063} (d)) is almost constant ($k \approx. 0.132$) with respect to $\langle n \rangle$, {\it i. e.} $v_1= 0.132 \Delta \mu$.
Since $p_1$ is also correlated with $|\Delta \mu|$ according to the relationship \cite{akutsu17}
\beq
p_1\approx  a+b |\Delta \mu|/\epsilon +c(|\Delta \mu|/\epsilon)^2, \label{eq_p1t063}
\eeq
where
$a=0.332$,  $b=15.6$, 
and $c=4.43\times 10^3$, 
we can solve Eq. (\ref{eq_p1t063}) with respect to $|\Delta \mu|/\epsilon$ to obtain
\beq
|\Delta \mu/\epsilon| = \frac{-b}{2c} + \sqrt{\frac{b^2}{4c^2}-\frac{a-p_1}{c}}.\label{eq_dmut063}
\eeq
Substituting Eq. (\ref{eq_p1lnn}) into Eq. (\ref{eq_dmut063}), we obtain
\beqa
|v_1|&=&(k\epsilon) |\Delta \mu|/\epsilon \nonumber \\
&=& 0.132 \times \left \{ -1.76 \times 10^{-3} \right. \nonumber \\
&& \hspace*{-5em} \left.+\sqrt{8.03\times 10^{-5} - 6.75\times 10^{-5}\ln \langle n \rangle} \right \}. \label{eq_v1-nt063}
\eeqa
The values of $v_1$ calculated from Eq. (\ref{eq_v1-nt063}) are shown by the solid lines in Fig. \ref{fig_p1-vstept063} (c), and these lines accurately reproduce the Monte Carlo data for $|\Delta \mu| < \Delta \mu_R$.
Therefore, the surface velocity can be expressed as
\beqa
|V|&=&  (k\epsilon) p_1 |\Delta \mu|/\epsilon \nonumber \\
&=& 0.132 \times (0.674 - 0.299 \ln \langle n \rangle)  \nonumber \\
&& \left \{ -1.76 \times 10^{-3} \right. \nonumber \\
&& \hspace*{-5em} \left.+\sqrt{8.03\times 10^{-5} - 6.75\times 10^{-5}\ln \langle n \rangle} \right \} \label{eq_v-nt063}
\eeqa
The lines obtained from Eq. (\ref{eq_v-nt063}) are indicated in Fig. \ref{V-n} (a) by solid lines, and also accurately reproduce the Monte Carlo data. 


\begin{figure*}
\centering
\includegraphics[width=12.0 cm,clip]{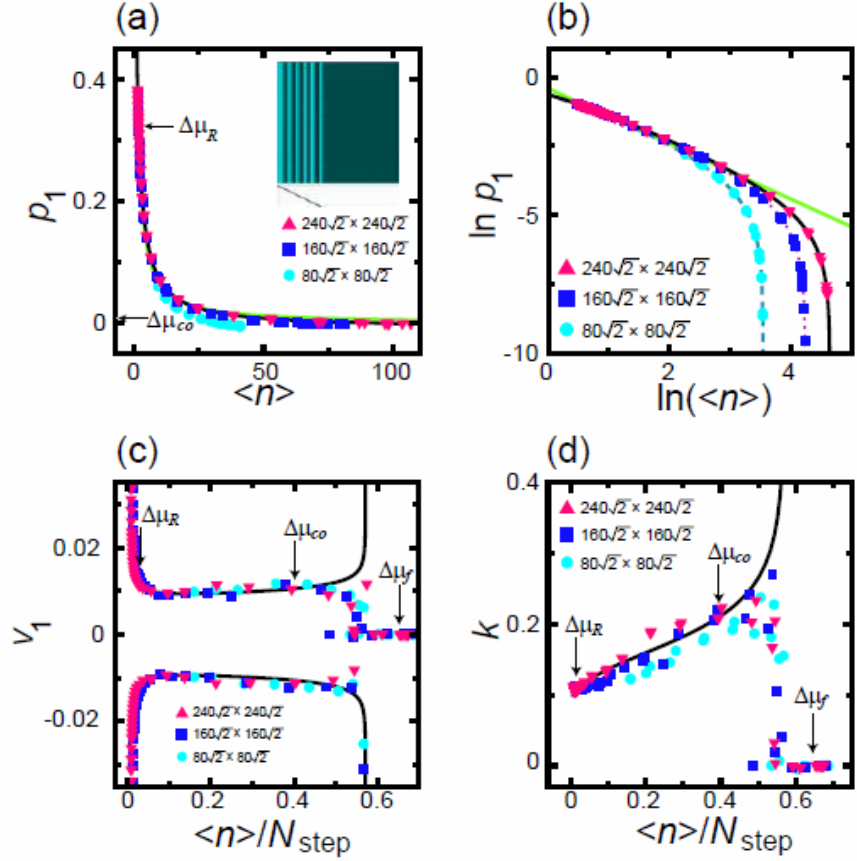}%
\caption{
(a) Macrostep height $\langle n \rangle$ dependence of the surface slope, $p_1$.
Symbols indicate $p_1$ values obtained using Eq. (\ref{eq_p1def}) based on the Monte Carlo data.
Solid line: Eq. (\ref{eq_p1def}) for $p_1$ with $N_{\rm step}=180$.
(b) $\ln p_1$ {\it v.s.} $\ln \langle n \rangle$.
Symbols indicate Monte Carlo data.
Solid line: Eq. (\ref{eq_p1def}) with $N_{\rm step}=180$.
Dotted line: Eq. (\ref{eq_p1def}) with $N_{\rm step}=120$.
Broken line: Eq. (\ref{eq_p1def}) with $N_{\rm step}=60$.
Light solid line: $\ln p_1 = -0.418- \ln \langle n \rangle$.
(c) $\langle n \rangle / N_{\rm step}$ dependence of the mean velocity of the elementary steps, $v_1$.
Symbols indicate values obtained using Eq. (\ref{eq_v1def}) with Monte Carlo data.
Dark solid line: Eq. (\ref{eq_k}) times $\Delta \mu$ with Eqs. (\ref{eq_p1_black}), (\ref{eq_n}) and $N_{\rm step}=180$.
(d) The $\langle n \rangle / N_{\rm step}$ dependence of kinetic coefficient $k$.
Symbols indicate values obtained using Eq. (\ref{eq_kdef}) with Monte Carlo data.
Solid line: Eq. (\ref{eq_k}) -- (\ref{eq_n}) with $N_{\rm step}=180$.
$\kBT/\epsilon=0.4$.
$\epsilon_{\rm int}/\epsilon=-0.9$. 
}  
\label{fig_p1vstep-nt04}
\end{figure*}

In addition, for $|\Delta \mu|> \Delta \mu_R$ (in which case only local macrosteps are present (Fig. \ref{fig_characteristics} (d))), and using $L= 240 \sqrt{2}$, we obtain
\beqa
&& \langle n \rangle - n_{\infty} = 9.16\times 10^{-6} (|\Delta \mu|/\epsilon)^{-\zeta}, \nonumber \\
&& \quad \zeta = 1.89 \pm 0.07, \\
&&|V| = 0.0855 (\Delta \mu)(|\Delta \mu|/\epsilon)^{\beta} , \nonumber \\
&& \quad \beta= 1.06 \pm 0.06.\label{eq_Vt063rough}
\eeqa
Therefore, we have
\beqa
|V| &=& 1.278\times 10^{-4} (\langle n \rangle-n_{\infty})^{-\beta/\zeta}, \nonumber \\
 && \ \beta/\zeta = 0.56 \pm 0.07, \label{eq_VndmuR} \\
|v_1| &=& 0.132 \times 2.16 \times 10^{-3} (\langle n \rangle - n_{\infty} )^{-1/\zeta}.
\label{eq_v1ndmuR}
\eeqa
The lines generated using Eqs. (\ref{eq_VndmuR}) and (\ref{eq_v1ndmuR}) are shown by the broken lines in Fig. \ref{V-n} (a) and Fig. \ref{fig_p1-vstept063} (c), respectively.
Although a large macrostep does not form, step velocity $v_1$ is lower than that produced by the original RSOS model because the locally merged steps pin the motion of the elementary steps.
As shown in Fig. \ref{fig_characteristics} (d), the elementary steps form a network of steps and, at the meeting points of neighboring steps, locally merged steps form local faceted macrosteps with finite lifetimes.
Because the side surface of the local faceted macrostep has few kinks, the growth of this region is quite slow \cite{akutsu12, akutsu14}.
In this manner, the local faceted macrostep pins the elementary steps.

\subsection{Step-faceting zone}

In the case of the step-faceting zone, the vicinal surface consists of (001) terrace surfaces and the (111) side surface (Fig. \ref{fig_nucleations}) at equilibrium, both of which are smooth.
The vicinal surface with the faceted macrostep in the non-equilibrium steady state shows all the configurations presented in Figs. \ref{fig_characteristics} (a) -- (d) depending on driving force $\Delta \mu$. 
The cases corresponding to Figs. \ref{fig_characteristics}(a) and (b) are explained in \S \ref{sec_vn}.
In this subsection, we analyze the data shown in Fig. \ref{fig_characteristics} (c).
If $\Delta \mu_{co} < |\Delta \mu| < \Delta \mu_R$, growth occurs via step-detachment/attachment, as in (Fig. \ref{fig_characteristics} (c)).
Since multi-nucleation occurs on both the (001) and (111) surfaces at the edge of the macrostep, the velocity of the macrostep is almost zero, and  surface velocity $V$ does not coincide with that of the macrostep.

In Fig. \ref{fig_p1vstep-nt04} (a), the $p_1$ values calculated from $\langle n \rangle$ data using Eq. (\ref{eq_p1def}) in conjunction with the Monte Carlo method are indicated by symbols, along with the $p_1$ values calculated using Eq. (\ref{eq_p1def}).  
These results demonstrates that $p_1$ falls to zero at $n^*$.
The line $p_1 \sim 1/\langle n \rangle$ is shown as a light green solid line in Fig. \ref{fig_p1vstep-nt04} (b).
This line reproduces the symbols accurately except for the data close to $\langle n \rangle \sim n^*$. 
In Figs. \ref{fig_p1vstep-nt04} (c), the $v_1$  values obtained by the Monte Carlo method in conjunction with Eq. (\ref{eq_v1def}) by using  $\langle n \rangle$ and $V$ values are indicated by symbols. 
In Figs. \ref{fig_p1vstep-nt04} (d), the $k$ values obtained by the Monte Carlo method in conjunction with Eq. (\ref{eq_kdef}) by using $\langle n \rangle$ and $V$ values are indicated by symbols.

The solid line in Fig. \ref{fig_p1vstep-nt04} (d) \cite{akutsu18} represents 
\beq
k =  a'+b' \exp[c'/|\Delta \mu/\epsilon|] , \label{eq_k}
\eeq
where $a'=0.094$, $b'=3.2 \times 10^{-3}$, and $c'=0.18$.
$|\Delta \mu|$ is related to $\langle n \rangle$ \cite{akutsu18} in association with $p_1$ according to 
\beqa
p_1&=& \frac{c_p}{\sqrt{|\Delta \mu/\epsilon|}} \exp \left[\frac{-g^*_p/2}{|\Delta \mu/\epsilon|-\Delta \mu_y(L)/\epsilon} \right],\nonumber \\
&&g^*_p = 0.423 \epsilon, \quad c_p=0.604,  \label{eq_p1_black} 
\eeqa
and
\beqa
\langle n \rangle &=&  \left [\frac{(\sqrt{2}-\bar{p})}{\bar{p}} \left ( \frac{\sqrt{2}}{p_1}-1 \right )^{-1} + \frac{1}{n^*}\right ]^{-1}. 
\label{eq_n}
\eeqa
Here, $\Delta \mu_y(L)$ is one of the characteristic driving forces (Table \ref{table1}),
$g_P^*$ is the total step free energy of the 2D critical island at the macrostep edge (calculated using the 2D Ising model) times $|\Delta \mu / \epsilon|$,
and $c_p$ is a coefficient related to the Zeldovich factor for the 2D nucleation.
The solid line obtained from Eq. (\ref{eq_k}) reproduces the Monte Carlo results well, except in the case that  $\langle n \rangle$ is close to $n^*$.
Here, $k$ increases as $\langle n \rangle$ increases due to the decreased step density on the terrace, {\it i. e.} $p_1 \rightarrow 0$.
In the case of a high step density on the terrace, the number of locally merged steps is also high.
Since each locally merged step is faceted and has few kinks, the velocity of the locally faceted macrostep is substantially lower than that of an elementary step.
As a result, the motions of steps are pinned randomly by the locally faceted macrosteps (Fig. \ref{fig_characteristics} (c)).
Conversely, when $\langle n \rangle$ is large, $|\Delta \mu|$ is small, the step density on the terrace is nearly zero, and the elementary steps meet neighboring steps less frequently.
Thus, the elementary steps travel like free steps and are not pinned as often.

It is interesting to observe that $v_1$ is almost constant except near $\langle n \rangle/N_{\rm step} \approx 0$ and $\langle n \rangle/N_{\rm step} \approx n^* / N_{\rm step}$. 
The solid line in Fig. \ref{fig_p1vstep-nt04} (c) was calculated using the equation $v_1=k \Delta \mu/\epsilon$, where $k$ is that in Eq. (\ref{eq_k}) and $\Delta \mu$ is related to the value of $\langle n \rangle$ by Eqs. (\ref{eq_p1_black}) and (\ref{eq_n}).
As $\langle n \rangle$ increases, $k$ increases but $|\Delta \mu|$ decreases, and so $v_1$ is almost unchanged over the approximate range of $\Delta \mu_{co} < |\Delta \mu| < \Delta \mu_R$.
Because $p_1 \propto 1/\langle n \rangle$, surface velocity $V$ becomes
\beq
V \propto 1/\langle n \rangle
\eeq 
over the approximate range of $\Delta \mu_{co} < |\Delta \mu| < \Delta \mu_R$, except for $\langle n \rangle \sim n^*$.
At  $\langle n \rangle \gtrsim n^*$, where $|\Delta \mu| \lesssim \Delta \mu_f$, there is one faceted macrostep and several elementary steps traveling on the terrace surface. 
In this case, because $V$ approaches zero more rapidly than does $p_1$ (due to the finite size effect), the $k$ values obtained from Eq. (\ref{eq_kdef}) are reduced to zero.

For $\Delta \mu_R < |\Delta \mu|$, the vicinal surface roughens kinetically and faceted macrosteps having finite lifetimes appear locally (Fig. \ref{fig_characteristics} (d)).
In the case that $L/a=240 \sqrt{2}$, $|\Delta \mu|$, $V$ and $v_1$ are expressed as
\beqa
&& |\Delta \mu|/\epsilon -\Delta \mu_{co}(L)/\epsilon
= 0.109 \ (\langle n \rangle - n_{\infty} )^{-1/\zeta'},\nonumber\\
&& \quad \zeta' = 1.57\pm 0.07, \quad n_{\infty}=1.33 \pm 0.08, \label{eq_dmu_rough}\\
&& |V| =  0.0677  \ (|\Delta \mu|/\epsilon -\Delta \mu_{co}(L)/\epsilon )^{\beta'},\nonumber\\
&& \quad \beta' = 1.19 \pm 0.05, \label{eq_V_rough}\\
&& v_1  = 0.442-0.489 |\Delta \mu| \label{eq_v1_rough},
\eeqa
where $\langle n \rangle$ is the mean height of the local macrosteps.
Substituting Eq. (\ref{eq_dmu_rough}) into Eqs. (\ref{eq_V_rough}) and (\ref{eq_v1_rough}) gives
\beqa
|V|&=&  4.87\times 10^{-3} (\langle n \rangle - n_{\infty} )^{-\beta'/\zeta'} \nonumber \\
&& \beta'/\zeta' = 0.76 \pm 0.06, \\
v_1  &=&  0.371 + 0.0602 \ (\langle n \rangle - n_{\infty} )^{-1/\zeta'}.
\eeqa
Similar to the results obtained for the step-droplet zone, the local macrosteps pin the step motion of the elementary steps.
Hence, the $k$ and $v_1$ values are both substantially smaller than those in the original RSOS model.

\section{Conclusions}
\begin{itemize}
\item The surface velocity $V$ decreases as the height of a faceted macrostep increases during reaction- (interface-) limited crystal growth, and this reduced velocity is related to a decrease in surface roughness or kink density.
\item In the step-droplet zone, the faceted macrostep grows (recedes) in the opposite direction to the growth (recession) direction of the elementary steps.
The decrease in the velocity of the faceted macrostep $V_m$ is also proportional to $\ln \langle n \rangle$ (Eq. (\ref{eq_v-nt063}), $|V_m| \propto |V|$).
\item In the step-droplet zone, the kinetic coefficient of the elementary steps is almost constant.
This coefficient has a value approximately one-third that of the kinetic coefficient of the original restricted solid-on-solid (RSOS) model because of the local faceted macrosteps on the terrace.   
\item In the step-faceting zone, the absolute value of the growth velocity of the surface, $|V|$, varies widely within the range defined by $|V| =$ constant $\times 1/\langle n \rangle$, because the height of the macrostep depends on the initial configuration of the vicinal surface.
\item In the step-faceting zone, the surface grows (recedes) via 2D multi-nucleation on both the (001) and (111) surfaces at the lower (upper) edge of the faceted macrostep. 
This nucleation on both surfaces inhibits the advancement of the faceted macrostep ($|V_m| \sim 0$).\\

\end{itemize}

\section{acknowledgments}
The author wishes to acknowledge helpful discussions with Prof. Y. Matsumoto, Prof. T. Nishinaga, Prof. T. Ujihara, Prof. K. Fujiwara, Prof. S. Uda and Prof. K. Kakimoto.
This work is supported by Kakenhi Grants-in-Aid 
(Nos. JP25400413 and JP17K05503) from the Japan Society for the Promotion of Science (JSPS).

\section{reference}
%


\section{End}

\end{document}